# A Sublogarithmic Approximation for Highway and Tollbooth Pricing

Iftah Gamzu[*]     Danny Segev[†]

**Abstract**

An instance of the tollbooth problem consists of an undirected network and a collection of single-minded customers, each of which is interested in purchasing a fixed path subject to an individual budget constraint. The objective is to assign a per-unit price to each edge in a way that maximizes the collective revenue obtained from all customers. The revenue generated by any customer is equal to the overall price of the edges in her desired path, when this cost falls within her budget; otherwise, that customer will not purchase any edge.

Our main result is a deterministic algorithm for the tollbooth problem on trees whose approximation ratio is $O(\log m / \log \log m)$, where $m$ denotes the number of edges in the underlying graph. This finding improves on the currently best performance guarantees for trees, due to Elbassioni et al. (SAGT '09), as well as for paths (commonly known as the highway problem), due to Balcan and Blum (EC '06). An additional interesting consequence is a computational separation between tollbooth pricing on trees and the original prototype problem of single-minded unlimited supply pricing, under a plausible hardness hypothesis due to Demaine et al. (SODA '06).

[*]Blavatnik School of Computer Science, Tel-Aviv University, Tel-Aviv 69978, Israel. Email: `iftgam@tau.ac.il`. Supported by the Israel Science Foundation, by the European Commission under the Integrated Project QAP funded by the IST directorate as Contract Number 015848, by a European Research Council (ERC) Starting Grant, and by the Wolfson Family Charitable Trust.

[†]Department of Statistics, University of Haifa, Haifa 31905, Israel. Email: `segevd@stat.haifa.ac.il`.

# 1 Introduction

An extensively-studied question in economics and operations management is that of pricing an assortment of products in a given market, trying to maximize revenue subject to a multitude of constraints. Somewhat informally, the inherent difficulty in such settings boils down to the obvious tension between two extremes: low prices attract more customers, while high prices generate greater revenues per purchase. Recently, the spotlights have been turned on computational challenges in pricing. What seems to be the driving force behind this line of research is an immense growth in the range of sources for acquiring costumer preferences data, which are now available as a result of the widespread use of the Internet.

**The tollbooth problem.** One particular computational task that has received much recent attention is the tollbooth problem, which captures optimization-related aspects of pricing connection links in networks, e.g., setting prices for road use in a system of toll highways. Formally, an instance of this problem consists of an undirected graph $G = (V, E)$ with $m$ edges, which can be broadly interpreted as products with unlimited supply. An additional ingredient of the input is a collection $C$ of $n$ single-minded customers, each of which is interested in purchasing a fixed path subject to an individual budget constraint. Technically speaking, the demand attributes of customer $i$ are represented by a pair $(P_i, b_i)$, where $P_i \subseteq G$ is the path she wishes to buy, and $b_i$ stands for her budget, namely, the maximum price she is willing to pay for that path. Any customer will buy a single unit of each edge in the desired path when its total cost falls within her budget; otherwise, she leaves without buying anything. With this setting in mind, the goal is to assign a per-unit price to each edge in a way that maximizes the overall revenue. More precisely, the objective is to compute a pricing scheme $p : E \to \mathbb{R}_+$ that maximizes the total revenue over all customers, $\sum_{i=1}^{n} R_p(i)$. Here, $R_p(i)$ denotes the revenue obtained from customer $i$, which evaluates to $\sum_{e \in P_i} p(e)$ when this cost does not exceed $b_i$, or to 0, otherwise.

**Previous work.** Guruswami et al. [23] seem to have been the first to study the tollbooth problem. Their main results in this context were to show that tollbooth pricing is APX-hard even when the underlying graph is a tree, and to devise exact dynamic-programming algorithms for the single-source variant on trees and for several other special cases. Additional hardness results were obtained by Briest and Krysta [7], who proved that even the seemingly-manageable setting of a simple path, commonly known as the highway problem, is in fact weakly NP-hard. Elbassioni, Raman, Ray, and Sitters [13] extended this result by establishing strong NP-hardness. On the positive side, however, Balcan and Blum [3] devised an $O(\log m)$ approximation for the highway problem; this finding is incomparable with the quasi-PTAS developed by Elbassioni, Sitters, and Zhang [14] later on. Finally, and very recently, Elbassioni et al. [13] proposed an $O(\log m)$ approximation for arbitrary trees. To conclude, approximating the tollbooth problem on trees, or even on simple paths, beyond the logarithmic threshold has remained an open research question.

## 1.1 Our results

The main result of this paper is a deterministic algorithm for the tollbooth problem on trees whose approximation ratio is $O(\log m / \log \log m)$, improving on the currently best performance guarantees for trees, as well as for paths, due to Elbassioni et al. [13], and to Balcan and Blum [3], respectively. Even though the quantitative magnitude of improvement is not that dramatic, our findings have additional interesting contributions:

**Conceptual.** We identify a computational separation between tollbooth pricing on trees and the original prototype problem of single-minded unlimited supply pricing (see Section 1.2). The latter cannot be approximated within a factor smaller than $\Omega(\log n)$, under a plausible hardness hypothesis regarding the balanced bipartite independent set problem [12], whereas in the former the resulting factor is better by $\Omega(\log \log n)$ or more. Regarding the relation between $m$ and $n$, we remark that an arbitrary instance of tollbooth pricing on trees can be reduced to one with $m = O(n)$. The general idea is that, when the number of edges is significantly larger than the number of customers, paths can be iteratively contracted (see, for instance, [13, Sec. 2.2]).

**Technical.** We believe that some of the algorithmic tools and analysis methods, illustrated in Sections 2 and 3, are of independent interest, and may be applicable in other settings as well. In particular, we have already derived new approximability bounds for graph orientation problems [18] by synthesizing ideas such as balanced decompositions, segment guessing, and randomization.

## 1.2 Related work

We proceed with a brief discussion on the single-minded unlimited supply pricing problem, from which the computational setting considered in this paper seems to have emerged. The input to the former problem consists of $m$ different products, with unlimited supply, and a collection of $n$ single-minded customers, each of which is interested in purchasing a particular subset (or bundle) of products subject to an individual budget constraint. The goal is to assign a per-unit price to each product in a way that maximizes the overall revenue, where again, a customer will buy a single unit of each product in her bundle only when its total cost fits within her budget. This problem was originally introduced by Guruswami et al. [23], who demonstrated that the single-price policy, where all products are given identical prices, guarantees an approximation ratio of $O(\log n + \log m)$ with respect to the optimal revenue. Later on, Balcan, Blum and Mansour [4] extended this finding to the setting of customers with general valuation functions. From a hardness point of view, Demaine, Feige, Hajiaghayi, and Salavatipour [12] established several inapproximability results under various complexity assumptions. In particular, they proved a lower bound of $\Omega(\log n)$, under a plausible hardness hypothesis regarding the balanced bipartite independent set problem.

A concurrent line of research focused on computing approximate pricing schemes in terms of other problem parameters. For instance, when the number of different products is fixed, Hartline and Koltun [24] showed that an FPTAS can be devised. Briest and Krysta [7] suggested an $O(\log B + \log \ell)$ approximation, where $B$ is the maximum number of requests per product and $\ell$ is the maximum size of any bundle, as well as a different algorithm, whose performance guarantee is $O(\ell^2)$. Balcan and Blum [3] improved on this result, to obtain a ratio of $O(\ell)$, and demonstrated that the vertex pricing problem (where $\ell = 2$) can be approximated within a factor of 4. Vertex pricing was further studied in [27, 26]. Recently, Cheung and Swamy [11] design an LP-based algorithm for a more general model of revenue maximization in a limited supply scenario that implies an $O(\log B)$ approximation for single-minded unlimited supply pricing.

We note that revenue maximization, in various colors and flavors, has received a great deal of recent attention in the computer science and operations research communities. Therefore, it is beyond the scope of this writing to do justice and present an exhaustive survey of previous work. We refer the reader to directly related papers [19, 16, 1, 5, 8, 9, 21, 6, 10, 22] and to the references therein for a more comprehensive review of the literature.

## 2 The Classification Process

Prior to describing the specifics of our approach in detail, which will inevitably involve delving into technicalities, it would be instructive to concentrate on the bigger picture. For this purpose, we begin by pointing out that the performance guarantee of $O(\log m / \log \log m)$ is obtained by employing the classify-and-select paradigm. More specifically, we exploit various structural properties to partition the collection of customers into $O(\log m / \log \log m)$ pairwise-disjoint classes. For each such class, given the additional structure imposed, we separately compute a pricing scheme whose overall revenue comes within a *constant factor* of the optimal revenue attainable from this class. In particular, each class is treated in a completely independent fashion, as if there are no other classes under consideration. Consequently, since the objective function is subadditive, the above-mentioned approximation ratio follows by picking, out of the set of all pricing schemes computed, the one that collects maximal revenue. We proceed by describing the customer classification process; this exposition will allow us to focus attention on the more involved single-class problem later on.

## 2.1 Classifying customers via balanced decompositions

In the following, we give a formal account of the process by which customers are partitioned into classes. To this end, we begin by introducing the notion of an almost-balanced decomposition, which can be viewed as a generalization of the well-known centroid decomposition [17]. It is worth noting that structural properties in this spirit have been explored and exploited in various settings (see, e.g., [15, 20, 25]).

**Definition 2.1.** Let $T = (V, E)$ be a tree. An *almost balanced $k$-decomposition* of $T$ is a partition of $T$ into $k$ edge-disjoint subtrees $T_1, \ldots, T_k$ such that each subtree contains between $|E|/(3k)$ and $3|E|/k$ edges.

For ease of presentation, we defer the proof of the following lemma to Appendix A.1.

**Lemma 2.2.** *Let $T = (V, E)$ be a tree with $|E| \geq k$. An almost balanced $k$-decomposition of $T$ exists and can be found in polynomial time. Moreover, the number of vertices that are shared by at least two subtrees is less than $k$.*

The classification process corresponds to a recursive decomposition of the input tree $T$; to better understand the upcoming discussion, we advise the reader to consult Figure 1. Let $\mathcal{T}_1 = \{T_1, \ldots, T_k\}$ be an almost balanced $k$-decomposition of $T$ into $k$ edge-disjoint subtrees. The first class of customers, $C_1$, consists of all customers separated by $\mathcal{T}_1$, that is, customers $i$ for which the endpoints of the desired path $P_i$ (henceforth, $s_i$ and $t_i$) reside in different subtrees of the decomposition $\mathcal{T}_1$. Now, to classify the remaining set of customers, $C \setminus C_1$, we recursively apply the previously-described procedure with respect to the collection of subtrees in $\mathcal{T}_1$. Specifically, in the second level of the recursion, an almost balanced $k$-decomposition is computed in each of the subtrees $T_1, \ldots, T_k$, to obtain a set $\mathcal{T}_2$, comprising of $k^2$ subtrees. The second class of customers, $C_2$, consists of all yet-unclassified customers separated by $\mathcal{T}_2$. In other words, the endpoints of each path $P_i$, for which $i \in C_2$, reside in different subtrees of $\mathcal{T}_2$, but in the same subtree of $\mathcal{T}_1$. The remaining classes $C_3, C_4, \ldots$ are defined similarly. It is important to note that the recursive process ends as soon as we arrive at a subtree with strictly less than $k$ edges. In this case, we make use of the trivial decomposition, where the given subtree is broken into individual edges.

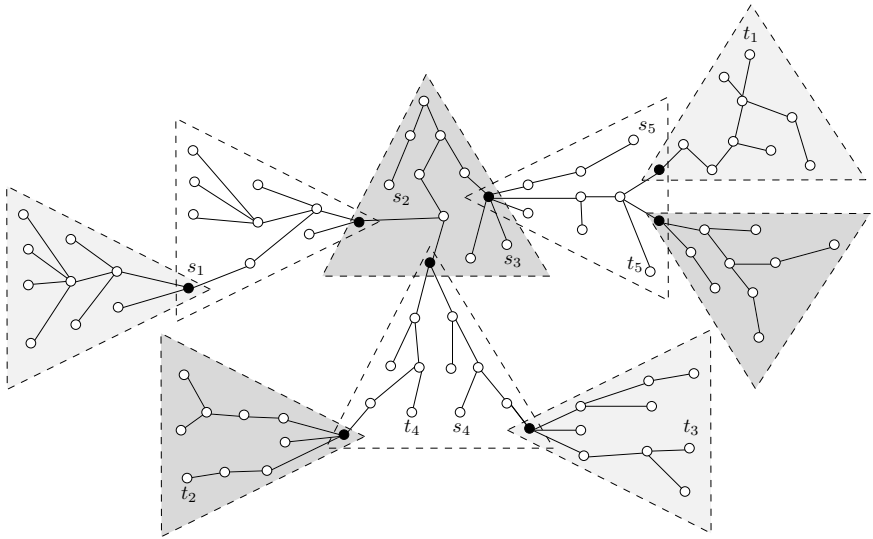

Figure 1: A schematic example for the collection of customers separated by an almost balanced decomposition, where each triangle marks a single subtree. Here, customers 1, 2, and 3 are separated, whereas 4 and 5 are not.

It is quite obvious that, up until this point in time, $k$ was treated as a parameter whose value has not been determined yet. For our purposes, we employ the above-mentioned classification process with $k = \lceil \log^{1/2} m \rceil$. With this value of $k$ at hand[1], one can easily verify that the overall number of levels in the recursion, or equivalently, the number of customer classes is $O(\log_k m) = O(\log m / \log \log m)$. This claim is immediately implied by observing that the maximum size of a subtree in level $\ell$ of the recursion is at most $(3/k)^\ell \cdot |E|$. As a side note, we remark that the balanced decomposition property, ensuring that all subtrees are of size roughly $|E|/k$, does not play any role from this point on, as its sole purpose was to restrict the number of classes to $O(\log m / \log \log m)$.

---

[1] A careful inspection later on shows that this particular selection is rather arbitrary, and any fixed exponent smaller than 1 will do the trick.

## 2.2 Why handling a single decomposition is sufficient

We remind the reader that a class of customers, say $C_\ell$, generally consists of several subsets of customers, each created when different subtrees in $\mathcal{T}_{\ell-1}$ are partitioned by the decomposition $\mathcal{T}_\ell$. More specifically, assuming that the subtrees in $\mathcal{T}_{\ell-1}$ are $T_1, T_2, \ldots$, the class $C_\ell$ can be written as the disjoint union of $C_\ell^1, C_\ell^2, \ldots$, where $C_\ell^j$ is the set of customers that are first separated when $T_j$ is partitioned. Notice that the desired path of any customer separated by some subtree decomposition must be contained in that subtree, since otherwise, this customer would have been separated in previous recursion steps. This observation implies that it is sufficient to compute an approximate pricing scheme for a single subtree decomposition and its induced set of separated customers. Given a polynomial-time algorithm that computes such pricing schemes, one can *sequentially* apply it to each of the subtree decompositions in the same recursion level. The resulting pricing schemes (in edge-disjoint subtrees) can then be "glued" to form a single scheme, defined for the entire edge set, collecting at least as much revenue as the sum of all individual subtree revenues.

## 3 The Single Decomposition Algorithm

In what follows, we focus our attention on a single decomposition, and devise a randomized algorithm that computes a pricing scheme whose expected revenue is within a constant factor of the optimal revenue attainable for this decomposition. Later on, we will argue that this algorithm can be easily derandomized. Formally, an instance of the problem in question consists of a tree $T = (V, E)$, and a partition $\mathcal{T} = \{T_1, \ldots, T_k\}$ of this tree into $k$ edge-disjoint subtrees, where $k \leq \lceil \log^{1/2} m \rceil$, such that the number of vertices shared by at least two subtrees is less than $k$. In addition, we are given a collection $C$ of $n$ customers, satisfying the following properties:

1. Each customer $i$ wishes to purchase a path $P_i$ so long as the overall price of this path does not exceed the budget $b_i$.

2. Each customer path $P_i$ is separated by $\mathcal{T}$, meaning that the endpoints of $P_i$ reside in different subtrees of the decomposition $\mathcal{T}$.

### 3.1 Notation and Terminology

For ease of presentation, it would be convenient to introduce some notation and terminology before laying down the nuts and bolts of our algorithm. To better understand the suggested notation, we refer the reader to a concrete example in Figure 2.

- Let $V_B \subseteq V$ be the set of *border vertices* of $\mathcal{T}$, that is, the set of vertices that are shared by at least two subtrees in $\mathcal{T}$. In addition, let $\mathcal{S} \subseteq T$ be the *skeleton* of $\mathcal{T}$, namely, the minimal subtree spanned by all border vertices. Note that this subtree consists of the union of paths connecting any two vertices in $V_B$.

- Let $p^* : E \to \mathbb{R}_+$ be an optimal pricing scheme, with an overall revenue of OPT.

- Now, recall that the endpoints of each customer path $P_i$ reside in different subtrees of the decomposition $\mathcal{T}$, meaning that $P_i$ must traverse at least one border vertex. Therefore, we can divide each customer path, with endpoints $s_i$ and $t_i$, into three (possibly empty) parts:

    1. A subpath between $s_i$ and its closest skeleton vertex $v_{s_i}$.

2. A subpath between $t_i$ and its closest skeleton vertex $v_{t_i}$.
3. A subpath between $v_{s_i}$ and $v_{t_i}$, along the skeleton.

Based on this definition, let $R^S_{p^*}(i)$, $R^T_{p^*}(i)$, and $R^M_{p^*}(i)$ denote the revenues obtained in the pricing scheme $p^*$ from the subpaths of customer $i$, respectively[2]. Clearly, $R_{p^*}(i) = R^S_{p^*}(i) + R^T_{p^*}(i) + R^M_{p^*}(i)$.

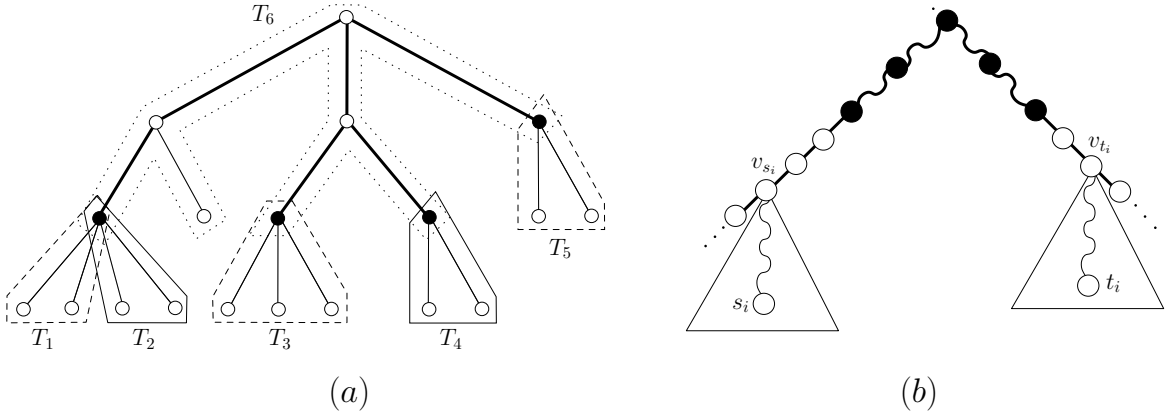

Figure 2: (a) An almost balanced 6-decomposition of a tree with 18 edges. Note that the black vertices are border vertices, and the heavy edges make up the skeleton $\mathcal{S}$. (b) Dividing the customer path $P_i$ into three parts.

### 3.2 The algorithm

Having all necessary definitions in place, we are now ready to present the specifics of our pricing algorithm, and to analyze its performance. In the remainder of this section, we consider two complementing scenarios, depending on the contribution of different path parts to the revenue generated by the optimal pricing scheme $p^*$. Somewhat informally, the first scenario captures a situation where a significant portion of OPT is gained from customer subpaths traversing non-skeleton edges, or in other words, when $\sum_{i=1}^{n}(R^S_{p^*}(i) + R^T_{p^*}(i)) = \Omega(1) \cdot \text{OPT}$. The second scenario corresponds to a situation where a large portion is delivered by subpaths along the skeleton, that is, when $\sum_{i=1}^{n} R^M_{p^*}(i) = \Omega(1) \cdot \text{OPT}$. For each of these scenarios, we compute a pricing scheme whose expected revenue is within a constant factor of optimal. As a result, we obtain a constant approximation ratio by computing both pricing schemes and picking the one that achieves the maximum revenue. For sake of simplicity, we begin by considering the easy-to-handle first scenario, noting that the second, more challenging scenario, will be discussed in the sequel.

**Scenario I:** $\sum_{i=1}^{n}(R^S_{p^*}(i) + R^T_{p^*}(i)) \geq \text{OPT}/2$

In the present setting, at least half of the optimal revenue is collected from customer subpaths consisting of non-skeleton edges, meaning that the collective contribution of the subpaths $P_i \setminus \mathcal{S}$, over all customers $i$, is at least OPT/2. The algorithmic tool that allows us to handle this scenario is a polynomial-time procedure, due to Guruswami et al. [23], for solving the single-source tollbooth problem on trees to optimality. Here, the underlying assumption is that all customer paths share a common endpoint, that is, $s_1 = \cdots = s_n$. With this tool at hand, the algorithm proceeds as follows:

1. Each of the decomposition subtrees, $T_1, \ldots, T_k$, is randomly and independently marked as being *active*, with probability $1/2$; otherwise, that tree is *inactive*.

2. We decide in advance that all skeleton edges, as well as all edges within inactive trees, could be purchased free of charge, i.e., their price is set to zero.

---
[2]These superscripts stand for: $S$ – subpath adjacent to $s_i$; $T$ – subpath adjacent to $t_i$; and $M$ – middle subpath.

3. Let $C(T_j)$ denote the subset of customers with an endpoint in $T_j$. For each active subtree $T_j$, we contract all skeleton edges in that subtree into a single root[3], designated by $r$, and apply the single-source algorithm, when the underlying set of customers are those in $C(T_j)$. Each such customer $i$ is now interested in purchasing the path connecting its endpoint in $T_j$ to the root $r$, still with a budget of $b_i$. Note that each customer has at most one endpoint in $T_j$, since otherwise, $T_j$ contains her desired path, in contradiction to the underlying setting.

**Lemma 3.1.** *The computed pricing scheme guarantees an expected revenue of at least* OPT/8.

**Proof.** We begin by observing that the expression $\sum_{i=1}^{n}(R_{p^*}^S(i) + R_{p^*}^T(i))$, which stands for the overall revenue in the optimal pricing scheme obtained from non-skeleton edges (over all customers), can be rewritten as the sum of revenues obtained from the subtrees $T_1, \ldots, T_k$. That is, recollecting that $C(T_j)$ denotes the subset of customers with an endpoint in $T_j$, we have

$$\sum_{i=1}^{n}\left(R_{p^*}^S(i) + R_{p^*}^T(i)\right) = \sum_{j=1}^{k} \sum_{i \in C(T_j)} R_{p^*}((P_i \setminus S) \cap T_j),$$

where $R_{p^*}((P_i \setminus S) \cap T_j)$ denotes the revenue obtained from the subpath of $P_i$ within the subtree $T_j$, excluding skeleton edges.

Now, for each active subtree $T_j$, one can easily verify that the pricing scheme $p^*$, restricted to the edges of $T_j$ (or, more precisely, to its skeleton-contracted version), constitutes a feasible solution to the corresponding single-source problem, with an objective value of at least $\sum_{i \in C(T_j)} R_{p^*}((P_i \setminus S) \cap T_j)$. Since for this particular problem we compute an optimal pricing scheme, the revenue collected from customers in $C(T_j)$, restricted to the subtree $T_j$, is at least that value. At this point in time, we observe that, for each such customer, with probability $1/2$ we will be able to stay within the budget $b_i$, and collect exactly the same profit for the entire path $P_i$, with additional edges outside of $T_j$. To establish this claim, since the price of all skeleton edges has been set to zero, it is sufficient to note that the subpath of $P_i$ that does not traverse edges either in the subtree $T_j$ or the skeleton $S$, i.e., $P_i \setminus (T_j \cup S)$, is entirely contained within a subtree different from $T_j$; that subtree will be marked as inactive with probability $1/2$, and its edge prices will be set to zero. It follows that the expected revenue of the pricing scheme computed is at least

$$\frac{1}{4} \sum_{j=1}^{k} \sum_{i \in C(T_j)} R_{p^*}((P_i \setminus S) \cap T_j) = \frac{1}{4} \sum_{i=1}^{n} \left(R_{p^*}^S(i) + R_{p^*}^T(i)\right) \geq \frac{\text{OPT}}{8}.$$

∎

### Scenario II: $\sum_{i=1}^{n} R_{p^*}^M(i) \geq \text{OPT}/2$

In this setting, at least half of the optimal revenue is collected from customer subpaths consisting of skeleton edges, meaning that the collective contribution of the subpaths $P_i \cap S$, over all customers $i$, is at least OPT/2. Our algorithm begins by first deciding that all non-skeleton edges could be purchased for free, and sets their price to zero. As a result, we may assume that the endpoints of each customer path are located on the skeleton, since otherwise, we can relocate them to their closest skeleton vertices, without any consequences whatsoever. With this structural alteration in mind, the skeleton pricing is carried out in two phases: *segment guessing*, where close estimates of the optimal prices along disjoint subpaths of the skeleton are obtained, followed by *randomized assignment*, where prices are associated with individual skeleton edges.

**Phase I: segment guessing.** We remind the reader that the skeleton $S$ is the minimal subtree of $T$ spanned by all border vertices $V_B$. We denote by $V_J$ the set of *junction vertices*, defined as non-border skeleton vertices with degree at least 3 (counting only skeleton edges). One can easily verify that $|V_J| <$ $|V_B| < k$. In addition, we call the vertex set $V_B \cup V_J$ the *core* of the skeleton $\mathcal{S}$. Based on these definitions, we can partition the skeleton into a collection $\Sigma(\mathcal{S})$ of edge-disjoint paths, which are referred to as *segments*. Each such segment is a subpath of $\mathcal{S}$ whose endpoints are core vertices, but its interior traverses only non-core vertices. Obviously, $|\Sigma(\mathcal{S})| = |V_B| + |V_J| - 1 < 2k$.

---

[3] The contraction is achieved by removing all skeleton edges, and unifying their endpoints into a representative vertex.

In what follows, we argue that one could obtain in polynomial time a close estimate for the total price $p^*(\sigma) = \sum_{e \in \sigma} p^*(e)$ of each segment $\sigma \in \Sigma(\mathcal{S})$, *simultaneously* for all segments. To this end, it is sufficient to prove that there exists a *very small* set of prices $\Gamma$, and a corresponding pricing scheme $p_\Gamma : E \to \mathbb{R}_+$, such that the price given by $p_\Gamma$ to every segment falls within $\Gamma$, namely, $p_\Gamma(\sigma) \in \Gamma$ for each $\sigma \in \Sigma(\mathcal{S})$, and such that $p_\Gamma(\sigma)$ approximates $p^*(\sigma)$ well. At the same time, we would like to make sure that the overall revenue that the supporting pricing scheme $p_\Gamma$ collects from customer subpaths consisting of skeleton edges still forms a fixed portion of $\sum_{i=1}^n R^M_{p^*}(i)$. A construction of this nature is given in the next lemma, whose proof appears in Appendix A.2.

**Lemma 3.2.** *Let $\Gamma = \{2^\ell \cdot b_{\max}/(4nm) : 0 \leq \ell \leq \lfloor \log(4nm^2) \rfloor\} \cup \{0\}$, where $b_{\max} = \max_i b_i$. Then, there is a pricing scheme $p_\Gamma : E \to \mathbb{R}_+$ that satisfies the following properties:*

1. *$p_\Gamma(\sigma) \in \Gamma$ for each $\sigma \in \Sigma(\mathcal{S})$.*

2. *$\sum_{i=1}^n R^M_{p_\Gamma}(i) \geq \sum_{i=1}^n R^M_{p^*}(i)/4$.*

By Lemma 3.2, we conclude that to obtain a close estimate for the total price $p^*(\sigma)$ of each segment $\sigma \in \Sigma(\mathcal{S})$, simultaneously for all segments, the total number of price assignments to be examined is of polynomial size, since

$$|\Gamma|^{|\Sigma(\mathcal{S})|} = (O(\log(nm)))^{O(k)} = (O(\log(nm)))^{O(\log^{1/2} m)} = o(nm) \ .$$

Consequently, we may assume without loss of generality that the set of segment prices $\{p_\Gamma(\sigma) : \sigma \in \Sigma(\mathcal{S})\}$, given in Lemma 3.2, is known in advance. This assumption can be easily enforced by enumerating over all $o(nm)$ possible assignments. On the other hand, it is worth noting that we do not assume any knowledge of the edge-specific pricing $p_\Gamma : E \to \mathbb{R}_+$.

**Phase II: randomized assignment.** The goal of this phase is to complete the pricing scheme by assigning carefully-picked random prices to individual skeleton edges, based on the estimated segment prices $\{p_\Gamma(\sigma) : \sigma \in \Sigma(\mathcal{S})\}$. The crux lies in making sure that the edge-specific prices *always* respect the outcome of the segment guessing phase, as stated in the next invariant.

**Invariant 3.3.** *With probability 1, the total price of each segment $\sigma \in \Sigma(\mathcal{S})$ is exactly $p_\Gamma(\sigma)$.*

Our assignment procedure consists of independent steps, where segments are processed one after the other. In each step, we consider a skeleton segment $\sigma = \langle v_1, \ldots, v_\ell \rangle$, and assign prices to its edges $(v_1, v_2), \ldots, (v_{\ell-1}, v_\ell)$ in a way that satisfies Invariant 3.3. Specifically, as illustrated in Figure 3(a), we pick one of the following four price assignments uniformly at random:

1. Assign a price of $p_\Gamma(\sigma)$ to the edge $(v_1, v_2)$, and zero prices to the remaining edges.

2. Assign a price of $p_\Gamma(\sigma)$ to the edge $(v_{\ell-1}, v_\ell)$, and zero prices to the remaining edges.

3. Assign prices based on a $v_1$-rooted single-source problem (see description below).

4. Assign prices based on a $v_\ell$-rooted single-source problem, analogous to item 3.

To complete the description of our algorithm, it remains to explain how the single-source instances in items 3 and 4 are created and solved; for brevity of presentation, we focus on the $v_1$-rooted case, noting that the opposite case is identical, up to changing the roles of $v_1$ and $v_\ell$. Once again, we will employ the dynamic-programming algorithm of Guruswami et al. [23] for solving the single-source tollbooth problem on trees, a valuable tool that was introduced in Scenario I. In particular, the $v_1$-rooted instance is comprised of the following components:

- The underlying graph is the segment $\sigma$.

- The set of customers are those with an endpoint in $\{v_2, \ldots, v_{\ell-1}\}$, whose desired path exits the segment $\sigma$ through $v_1$.

- For each customer $i$ under consideration, we set up a new endpoint at $v_1$ instead of the one outside $\sigma$, and change her budget to $\min\{p_\Gamma(\sigma), b_i - \sum_{\bar\sigma : \bar\sigma \subseteq P_i} p_\Gamma(\bar\sigma)\}$. Note that the latter term is exactly the budget remaining to purchase $P_i$, assuming that a total price of $p_\Gamma(\bar\sigma)$ has already been paid for each segment $\bar\sigma$ that is fully-contained in $P_i$.

Needless to say, the single-source algorithm does not set a price for the edge $(v_{\ell-1}, v_\ell)$, as it is not contained in any desired path of the $v_1$-rooted instance constructed. Therefore, to ensure that Invariant 3.3 holds, we set the price of $(v_{\ell-1}, v_\ell)$ to be the difference between the total segment price $p_\Gamma(\sigma)$ and the total newly-computed price of $(v_1, v_2), \ldots, (v_{\ell-2}, v_{\ell-1})$. Here, it is important to point out that this difference is indeed non-negative, since the dynamic-programming algorithm guarantees that the price of each subpath $\langle v_1, \ldots, v_j \rangle$ is equal to the budget of some customer whose desired path is contained in $\langle v_1, \ldots, v_j \rangle$ (see [23, Thm. 5.3]); on the other hand, the budget of every customer cannot exceed $p_\Gamma(\sigma)$, by definition.

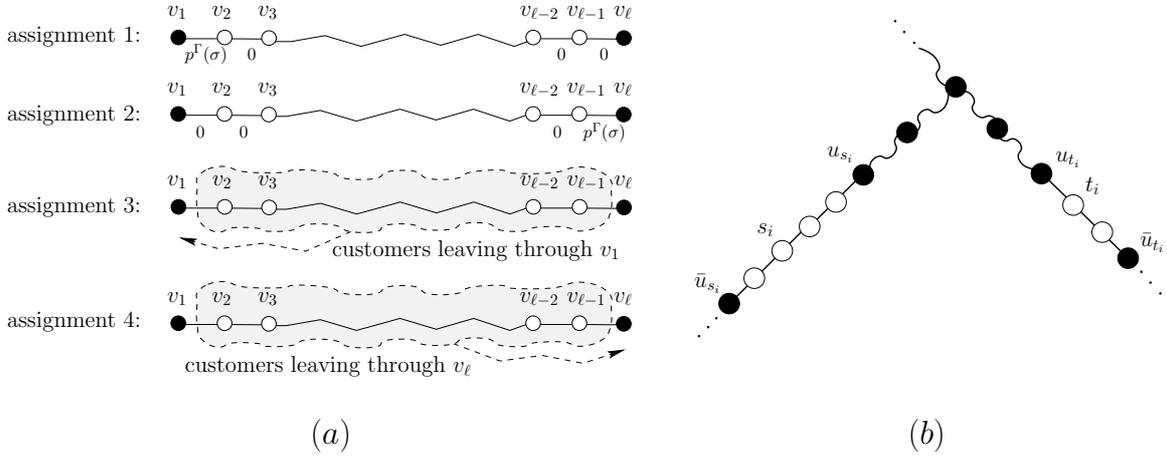

Figure 3: (a) A schematic description of the four random assignments for the segment $\sigma$. Here, core vertices are marked in black. (b) The new customer path partition.

**Analysis.** The remainder of this section is devoted to proving that the expected revenue of the pricing scheme computed in the randomized assignment phase is within a constant factor of optimal, as formally stated in the following lemma.

**Lemma 3.4.** *The pricing scheme constructed in the randomized assignment phase guarantees an expected revenue of at least* OPT/256.

Recall that we have previously assumed the endpoints of each customer path $P_i$ to be located on the skeleton. Moreover, since these endpoints reside in different subtrees of the decomposition $\mathcal{T}$, the path $P_i$ must traverse at least one border vertex. For this reason, as shown in Figure 3(b), we can divide each customer path, with endpoints $s_i$ and $t_i$, into three (possibly empty) parts:

1. A subpath, along a partial segment, between $s_i$ and its closest core vertex $u_{s_i}$.

2. A subpath, along a partial segment, between $t_i$ and its closest core vertex $u_{t_i}$.

3. A subpath between $u_{s_i}$ and $u_{t_i}$, along a sequence of complete segments.

With these definitions in mind, let $R_{p_\Gamma}^{\tilde{S}}(i)$, $R_{p_\Gamma}^{\tilde{T}}(i)$, and $R_{p_\Gamma}^{\tilde{M}}(i)$ denote the revenues obtained in the estimated pricing scheme $p_\Gamma$ from the subpaths of customer $i$, respectively. We remark that $p_\Gamma$ is the pricing scheme whose existence has been established in Lemma 3.2. Note that, by item 2 of this lemma, we can bound the sum of the above-mentioned revenues, over all customers, in terms of OPT, since

$$\sum_{i=1}^n \left( R_{p_\Gamma}^{\tilde{S}}(i) + R_{p_\Gamma}^{\tilde{T}}(i) + R_{p_\Gamma}^{\tilde{M}}(i) \right) = \sum_{i=1}^n R_{p_\Gamma}^M(i) \geq \frac{1}{4} \sum_{i=1}^n R_{p^*}^M(i) \geq \frac{\text{OPT}}{8} \ .$$

In what follows, we consider two cases, depending on the contribution of different path parts to the revenue generated by the pricing scheme $p_\Gamma$. For each of these cases, we show that the randomized assignment phase computes a pricing scheme whose expected revenue is within a constant factor of optimal.

**Case I:** $\sum_{i=1}^n R_{p_\Gamma}^{\tilde{M}}(i) \geq \text{OPT}/16$. In this setting, a significant fraction of the optimal revenue is collected from customer subpaths consisting of complete segments, meaning that the collective contribution of the paths $u_{s_i} \leftrightsquigarrow u_{t_i}$, over all customers $i$, is at least OPT/16. Notice that, by Invariant 3.3, our pricing scheme always assigns a total price of $p_\Gamma(\sigma)$ to each segment $\sigma \in \Sigma(S)$, implying that the overall price of $u_{s_i} \leftrightsquigarrow u_{t_i}$ is exactly $R_{p_\Gamma}^{\tilde{M}}(i)$. Therefore, the revenue from customer $i$ is at least $R_{p_\Gamma}^{\tilde{M}}(i)$, unless the total price of the remaining partial segments $s_i \leftrightsquigarrow u_{s_i}$ and $t_i \leftrightsquigarrow u_{t_i}$ exceeds the residual budget, $b_i - R_{p_\Gamma}^{\tilde{M}}(i)$. We next argue that, with probability at least 1/16, all edges on these partial segments will be given zero prices. Consequently, the expected revenue from customer $i$ is at least $R_{p_\Gamma}^{\tilde{M}}(i)/16$, and by linearity of expectation, the overall expected revenue is no less than $\sum_{i=1}^n R_{p_\Gamma}^{\tilde{M}}(i)/16 \geq \text{OPT}/256$.

For the purpose of establishing the previously mentioned claim, we will show that, with probability at least 1/4, each and every edge on $s_i \leftrightsquigarrow u_{s_i}$ is assigned a zero price. The claim then follows by observing that an identical property with respect to $t_i \leftrightsquigarrow u_{t_i}$ can be proven along the same lines, and also, that the two events are independent. Now, notice that if $s_i$ is a core vertex then the claim trivially holds, since $s_i \leftrightsquigarrow u_{s_i}$ is empty. Hence, let us consider the case where $s_i$ is an internal vertex on the segment between the core vertices $u_{s_i}$ and $\bar{u}_{s_i}$. In this case, with probability 1/4, our algorithm assigns a price of zero to all edges in this segment, except for the edge adjacent to $\bar{u}_{s_i}$; in particular, all edges on $s_i \leftrightsquigarrow u_{s_i}$ are assigned zero prices.

**Case II:** $\sum_{i=1}^n (R_{p_\Gamma}^{\tilde{S}}(i) + R_{p_\Gamma}^{\tilde{T}}(i)) \geq \text{OPT}/16$. Here, a constant fraction of the optimal revenue is collected from customer subpaths consisting of partial segments, meaning that the collective contribution of the paths $s_i \leftrightsquigarrow u_{s_i}$ and $t_i \leftrightsquigarrow u_{t_i}$, over all customers $i$, is at least OPT/16. Consider some segment $\sigma = \langle v_1, \ldots, v_\ell \rangle$, and let $C(\sigma)$ denote the collection of customers that have one of their path endpoints in $\{v_2, \ldots, v_{\ell-1}\}$. Clearly, if we use $R_{p_\Gamma}^\sigma(i)$ to denote the revenue obtained in the pricing scheme $p_\Gamma$ from customer $i$ along the segment $\sigma$, then

$$\sum_{\sigma \in \Sigma(S)} \sum_{i \in C(\sigma)} R_{p_\Gamma}^\sigma(i) = \sum_{i=1}^n \left( R_{p_\Gamma}^{\tilde{S}}(i) + R_{p_\Gamma}^{\tilde{T}}(i) \right) \ .$$

Furthermore, let $C^L(\sigma)$ and $C^R(\sigma)$ be the sets of customers in $C(\sigma)$ whose desired path exits the segment $\sigma$ through $v_1$ and $v_\ell$, respectively. Obviously, $C(\sigma) = C^L(\sigma) \cup C^R(\sigma)$.

Recall that, with probability 1/4, our algorithm computes an optimal pricing scheme $p_L$ for a $v_1$-rooted single-source problem on the segment $\sigma$. In particular, it is not difficult to verify that the underlying set of customers would be $C^L(\sigma)$, with paths restricted to the segment $\sigma$, and moreover, that each customer $i$ will be associated with a newly-defined budget of $\min\{p_\Gamma(\sigma), b_i - \sum_{\bar{\sigma}: \bar{\sigma} \subseteq P_i} p_\Gamma(\bar{\sigma})\}$. Notice that the pricing scheme $p_\Gamma$ forms a feasible solution to this instance, implying that if $R_{p_L}^\sigma(i)$ denotes the overall revenue from customer $i$ along the segment $\sigma$ with respect to $p_L$, we have

$$\sum_{i \in C^L(\sigma)} R_{p_L}^\sigma(i) \geq \sum_{\substack{i \in C^L(\sigma): \\ s_i \in \sigma}} R_{p_\Gamma}^{\tilde{S}}(i) + \sum_{\substack{i \in C^L(\sigma): \\ t_i \in \sigma}} R_{p_\Gamma}^{\tilde{T}}(i) \ .$$

Similarly, with probability 1/4, we compute an optimal pricing scheme $p_R$ for a $v_\ell$-rooted single-source problem, and using almost identical arguments, obtain the inequality

$$\sum_{i \in C^R(\sigma)} R^\sigma_{p_R}(i) \geq \sum_{\substack{i \in C^R(\sigma): \\ s_i \in \sigma}} R^{\tilde{S}}_{p_\Gamma}(i) + \sum_{\substack{i \in C^R(\sigma): \\ t_i \in \sigma}} R^{\tilde{T}}_{p_\Gamma}(i) \ .$$

We proceed by showing that for every productive customer $i \in C(\sigma)$, with probability at least 1/4, the total price that will be assigned to the path $P_i$, excluding edges residing within the segment $\sigma$, is exactly $\sum_{\bar{\sigma}: \bar{\sigma} \subseteq P_i} p_\Gamma(\bar{\sigma})$. In this case, the (global) budget $b_i$ is not exceeded, since customer $i$ was given a budget of no more than $b_i - \sum_{\bar{\sigma}: \bar{\sigma} \subseteq P_i} p_\Gamma(\bar{\sigma})$ in the single-source problem on $\sigma$. Prior to proving this claim, we remark that it is indeed sufficient to conclude Case II, since the overall expected revenue would be at least

$$\frac{1}{16} \sum_{\sigma \in \Sigma(S)} \left( \sum_{i \in C^L(\sigma)} R^\sigma_{p_L}(i) + \sum_{i \in C^R(\sigma)} R^\sigma_{p_R}(i) \right)$$

$$\geq \frac{1}{16} \sum_{\sigma \in \Sigma(S)} \left( \sum_{\substack{i \in C^L(\sigma): \\ s_i \in \sigma}} R^{\tilde{S}}_{p_\Gamma}(i) + \sum_{\substack{i \in C^L(\sigma): \\ t_i \in \sigma}} R^{\tilde{T}}_{p_\Gamma}(i) + \sum_{\substack{i \in C^R(\sigma): \\ s_i \in \sigma}} R^{\tilde{S}}_{p_\Gamma}(i) + \sum_{\substack{i \in C^R(\sigma): \\ t_i \in \sigma}} R^{\tilde{T}}_{p_\Gamma}(i) \right)$$

$$= \frac{1}{16} \sum_{\sigma \in \Sigma(S)} \sum_{i \in C(\sigma)} \left( R^{\tilde{S}}_{p_\Gamma}(i) + R^{\tilde{T}}_{p_\Gamma}(i) \right)$$

$$= \frac{1}{16} \sum_{i=1}^{n} \left( R^{\tilde{S}}_{p_\Gamma}(i) + R^{\tilde{T}}_{p_\Gamma}(i) \right)$$

$$\geq \frac{\text{OPT}}{256} \ .$$

To prove the last claim, consider some particular customer $i \in C(\sigma)$, and without loss of generality, suppose that the endpoint of $P_i$ that resides within the segment $\sigma$ is $s_i$. We therefore focus on bounding the total price of the subpath connecting $u_{s_i}$ and $t_i$, which can be broken into two probabilistically-independent parts:

- A subpath between $u_{s_i}$ and $u_{t_i}$, along a sequence of complete segments. By Invariant 3.3, with probability 1, the total price of each complete segment $\bar{\sigma}$ is $p_\Gamma(\bar{\sigma})$, immediately implying that the total price of $u_{s_i} \leftrightsquigarrow u_{t_i}$ is $\sum_{\bar{\sigma}: \bar{\sigma} \subseteq P_i} p_\Gamma(\bar{\sigma})$.

- A subpath, along a partial segment (different from $\sigma$), between $t_i$ and $u_{t_i}$. Here, with probability at least 1/4, each and every edge on $t_i \leftrightsquigarrow u_{t_i}$ is assigned a zero price. The arguments for proving this case are identical to those in Case I, and we do not repeat them to avoid duplicity.

### 3.3 Derandomization

The avid reader may already have noticed that the extent to which we utilize randomization is rather limited, and that its foremost purpose is to make the presentation of our algorithm significantly simpler. More specifically, in Scenario I each subtree in the decomposition is randomly marked as being active or inactive, whereas in Scenario II one of four possible price assignments is picked at random for each segment. In other words, all we need to obtain a deterministic algorithm are two uniform sample spaces, with $O(1)$ possible values for $O(\log^{1/2} m)$ random variables. These can be constructed in polynomial time either explicitly, as there are only $O(m^{O(1)})$ outcomes to examine, or in a more compact way, by observing that nothing more than pairwise-independence (see, for instance, [2, Chap. 15]) is required for the preceding analysis.

## A  Additional Proofs

### A.1  Proof of Lemma 2.2

We begin by presenting a well-known result regarding centroid decompositions in trees.

**Definition A.1.** Let $T = (V, E)$ be a tree. A *centroid decomposition* of $T$ is a partition of $T$ into two edge-disjoint subtrees (sharing a common vertex) such that each subtree contains between $|E|/3$ and $2|E|/3$ edges.

**Theorem A.2.** ([17]) *Let $T = (V, E)$ be a tree with $|E| \geq 2$. A centroid decomposition of $T$ exists and can be found in linear time.*

We now propose an iterative process for generating an almost balanced $k$-decomposition of $T$ in polynomial time. This process consists of $k - 1$ steps, where in each step a centroid decomposition is applied to a subtree having maximal number of edges. That is, in the first step, a centroid decomposition is applied to $T$ and results in subtrees $T_1$ and $T_2$; in the second step, a centroid decomposition is applied to the subtree with maximal number of edges (out of $T_1$ and $T_2$); so forth and so on. Clearly, this process can be implemented to run in polynomial time, as implied by Theorem A.2.

We argue that the collection of $k$ subtrees resulting from this process is indeed an almost balanced $k$-decomposition, noting in advance that the number of vertices shared by at least two subtrees is less than $k$. For this purpose, we are left to prove that each subtree contains between $|E|/(3k)$ and $3|E|/k$ edges. We next claim that in each step, the number of edges in the maximal subtree, i.e., the subtree that has maximal number of edges, is no more than 3 times greater than the number of edges in the minimal subtree, i.e., the subtree with minimal number of edges. This claim implies that, at the end of the process, each subtree must have at least $|E|/(3k)$ edges. Otherwise, it follows that all $k$ subtrees must have less than $|E|/k$ edges, and thus, the decomposition consists of less than $|E|$ edges. By employing similar arguments, one can show that each subtree has no more than $3|E|/k$ edges.

We turn to prove the above-mentioned claim by induction on the number of steps performed. At the beginning of the process, the claim trivially holds. Now, suppose the claim holds at the beginning of some step. Namely, the maximal subtree has $r_{\max}$ edges, the minimal subtree has $r_{\min}$ edges, and $r_{\max}/r_{\min} \leq 3$. Recall that the centroid decomposition partitions a subtree with maximal number of edges into two subtrees that contain between $r_{\max}/3$ and $2r_{\max}/3$ edges. This implies that at the beginning of the following step, the maximal subtree has no more than $r_{\max}$ edges, while the minimal subtree has at least $\min\{r_{\min}, r_{\max}/3\}$ edges. Notice that in either case, the size ratio between the maximal and minimal subtrees is at most 3.

## A.2 Proof of Lemma 3.2

The general idea behind our proof is to perform a sequence of modifications to the optimal pricing scheme $p^*$, trying to arrive at a new scheme $p_\Gamma$ that satisfies the required properties. For this purpose, we begin by observing that, without loss of generality, $p^*(e) \leq b_{\max}$ for every edge $e \in E$; otherwise, by setting $p^*(e) = b_{\max}$, the overall revenue may only improve. We now separately consider each segment $\sigma \in \Sigma(\mathcal{S})$, and proceed as follows:

1. If $p^*(\sigma) < b_{\max}/(4nm)$, we set the price of every edge in $\sigma$ to zero.

2. If $b_{\max}/(4nm) \leq p^*(\sigma) \leq mb_{\max}$, we uniformly scale down the prices of all edges in $\sigma$ such that the newly-defined price of $\sigma$ will be equal to the maximal value $\gamma \in \Gamma$ such that $\gamma \leq p^*(\sigma)$. For this purpose, the scaling factor is simply $\gamma/p^*(\sigma) \geq 1/2$, where the last inequality holds since $\Gamma$ consists of a geometric sequence between $b_{\max}/(4nm)$ and $mb_{\max}$, with a multiplier of 2.

This construction clearly satisfies the first property, and it remains to prove $\sum_{i=1}^n R^M_{p_\Gamma}(i) \geq \sum_{i=1}^n R^M_{p^*}(i)/4$. To validate the last inequality, note that our modifications may lead to two types of revenue losses: additive (due to item 1) and multiplicative (due to item 2). The additive loss can be easily bounded by observing that each edge appears at most once on any customer path and that the skeleton consists of at most $m$ edges. Therefore, the entire contribution of edges that were modified in the first item can be bounded by

$$nm \cdot \frac{b_{\max}}{4nm} = \frac{b_{\max}}{4} \leq \frac{\text{OPT}}{4} \leq \frac{1}{2} \sum_{i=1}^n R^M_{p^*}(i) \, ,$$

where the first inequality holds since $\text{OPT} \geq b_{\max}$, as an overall revenue of $b_{\max}$ can obviously be attained. In addition, the multiplicative loss can be bounded by noting that the scaling factor in the second item is at least $1/2$. It follows that $\sum_{i=1}^n R^M_{p_\Gamma}(i) \geq \sum_{i=1}^n R^M_{p^*}(i)/4$.